\documentclass[twocolumn,aps,showpacs]{revtex4}
\usepackage{amsmath}
\usepackage{amssymb}
\usepackage{epsf}

\input{tcilatex}

\begin{document}

\title{Weak localisation magnetoresistance and valley symmetry in graphene}
\author{E. McCann$^{1}$, K. Kechedzhi$^{1}$, Vladimir I. Fal'ko$^{1}$, H.
Suzuura$^{2}$, T. Ando$^{3}$, and B.L. Altshuler$^{4}$}
\affiliation{$^{1}$Department of Physics, Lancaster University, Lancaster, LA1 4YB, UK \\
$^{2}$Division of Applied Physics, Graduate School of Engineering, Hokkaido
University, Sapporo 060-8628, Japan\\
$^{3}$Department of Physics, Tokyo Institute of Technology, 2-12-1 Ookayama,
Meguro-ku, Tokyo 152-8551, Japan\\
$^{4}$Physics Department, Columbia University, 538 West 120th Street, New
York, NY 10027}

\begin{abstract}
Due to the chiral nature of electrons in a monolayer of graphite (graphene)
one can expect weak antilocalisation and a positive weak-field
magnetoresistance in it. However, trigonal warping (which breaks $\mathbf{p}%
\rightarrow -\mathbf{p}$ symmetry of the Fermi line in each valley)
suppresses antilocalisation, while inter-valley scattering due to atomically
sharp scatterers in a realistic graphene sheet or by edges in a narrow wire
tends to restore conventional negative magnetoresistance. We show this by
evaluating the dependence of the magnetoresistance of graphene on relaxation
rates associated with various possible ways of breaking a 'hidden' valley
symmetry of the system.
\end{abstract}

\maketitle

The chiral nature \cite{AndoQHE,AndoNoBS,McCann,CheianovFalko} of
quasiparticles in graphene (monolayer of graphite), which originates from
its honeycomb lattice structure and is revealed in quantum Hall effect
measurements \cite{Novoselov,Zhang}, is attracting a lot of interest. In
recently developed graphene-based transistors \cite{Novoselov,Zhang} the
electronic Fermi line consists of two tiny circles \cite{AndoReview}
surrounding corners $\mathbf{K}_{\pm }$ of the hexagonal Brillouin zone \cite%
{kpoints}, and quasiparticles are described by 4-component Bloch functions $%
\Phi =$[$\phi _{\mathbf{K}_{+},A}$, $\phi _{\mathbf{K}_{+},B}$, $\phi _{%
\mathbf{K}_{-,}B}$, $\phi _{\mathbf{K}_{-},A}$], which characterise
electronic amplitudes on two crystalline sublattices ($A$ and $B$), and the
Hamiltonian 
\begin{equation}
{\hat{H}}=v\,\Pi _{z}\otimes \mathbf{\sigma p}-\mu \lbrack \sigma _{x}({p}%
_{x}^{2}-{p}_{y}^{2})-2\sigma _{y}{p}_{x}{p}_{y}].  \label{h1}
\end{equation}%
Here, we use direct products of Pauli matrices $\sigma _{x,y,z},\sigma
_{0}\equiv \hat{1}$ acting in the sublattice space ($A,B$) and $\Pi
_{x,y,z},\Pi _{0}\equiv \hat{1}$ acting in the valley space ($\mathbf{K}%
_{\pm }$) to highlight the form of ${\hat{H}}$ in the non-equivalent valleys 
\cite{kpoints}. Near the center of each valley electron dispersion is
determined by the Dirac-type part $v\,\mathbf{\sigma p}$ of ${\hat{H}}$. It
is isotropic and linear. For the valley $\mathbf{K}_{+}$ the electronic
excitations with momentum \textbf{p} have energy $vp$ and are chiral with $%
\mathbf{\sigma p}/p=1$, while for holes the energy is $-vp$ and $\mathbf{%
\sigma p}/p=-1$. In the valley $\mathbf{K}_{-}$, the chirality is inverted:
it is $\mathbf{\sigma p}/p=-1$ for electrons and $\mathbf{\sigma \!p}/p=1$
for holes. The quadratic term in Eq. (\ref{h1}) violates the isotropy of the
Dirac spectrum and causes a weak trigonal warping \cite{kpoints}.

Due to the chirality of electrons in a graphene-based transistor, charges
trapped in the substrate or on its surface cannot scatter carriers in
exactly the backwards direction \cite{AndoNoBS,AndoReview}, provided that
they are remote from the graphene sheet by more than the lattice constant.
In the theory of quantum transport \cite{WL} the suppression of
backscattering is associated with weak anti-localisation (WAL) \cite%
{LarkinWAL}. For purely potential scattering, possible WAL in graphene has
recently been related to the Berry phase $\pi $ specific to the Dirac
fermions, though it has also been noticed that conventional weak
localisation (WL) may be restored by intervalley scattering \cite%
{AndoWL,khve06}.

In this Letter we show that the WL magnetoresistance in graphene directly
reflects the degree of valley symmetry breaking by the warping term in the
free-electron Hamiltonian (\ref{h1}) and by atomically sharp disorder. To
describe the valley symmetry, we introduce two sets of 4$\times $4 Hermitian
matrices: 'isospin' $\vec{\Sigma}=(\Sigma _{x},\Sigma _{y},\Sigma _{z})$ and
'pseudospin' $\vec{\Lambda}=(\Lambda _{x},\Lambda _{y},\Lambda _{z})$. These
are defined as 
\begin{eqnarray}
\Sigma _{x} &=&\Pi _{z}\otimes \sigma _{x},\;\Sigma _{y}=\Pi _{z}\otimes
\sigma _{y},\;\Sigma _{z}=\Pi _{0}\otimes \sigma _{z},  \label{Sigma} \\
\Lambda _{x} &=&\Pi _{x}\otimes \sigma _{z},\;\Lambda _{y}=\Pi _{y}\otimes
\sigma _{z},\;\Lambda _{z}=\Pi _{z}\otimes \sigma _{0},  \label{Lamda}
\end{eqnarray}%
and form two mutually independent algebras, $[\vec{\Sigma},\vec{\Lambda}]=0$%
, 
\begin{equation*}
\lbrack \Sigma _{s_{1}},\Sigma _{s_{2}}]=2i\varepsilon ^{s_{1}s_{2}s}\Sigma
_{s},\;[\Lambda _{l_{1}},\Lambda _{l_{2}}]=2i\varepsilon
^{l_{1}l_{2}l}\Lambda _{l},
\end{equation*}%
which determine two commuting subgroups of the group U$_{4}$ of unitary
transformations \cite{U4} of a 4-component $\Phi $: an isospin (sublattice)
group SU$_{2}^{\Sigma }\equiv \{e^{ia\vec{n}\cdot \!\vec{\Sigma}}\}$ and a
pseudospin (valley) group SU$_{2}^{\Lambda }\equiv \{\mathrm{e}^{ib\vec{n}%
\cdot \!\vec{\Lambda}}\}$.

The operators $\vec{\Sigma}$ and $\vec{\Lambda}$ help us to represent the
electron Hamiltonian in weakly disordered graphene as 
\begin{gather}
{\hat{H}}=v\,\vec{\Sigma}\mathbf{p}+{\hat{h}}_{\mathrm{w}}+\mathrm{\hat{I}}u(%
\mathbf{r})+\sum_{s,l=x,y,z}\Sigma _{s}\Lambda _{l}u_{s,l}(\mathbf{r}),
\label{h1-2} \\
\mathrm{where}\;\;{\hat{h}}_{\mathrm{w}}=-\mu \Sigma _{x}(\,\vec{\Sigma}%
\mathbf{p})\Lambda _{z}\Sigma _{x}(\,\vec{\Sigma}\mathbf{p})\Sigma _{x}. 
\notag
\end{gather}%
The Dirac part of ${\hat{H}}$ in Eq.(\ref{h1-2}), $v\,\vec{\Sigma}\mathbf{p}$
and potential disorder $\mathrm{\hat{I}}u(\mathbf{r})$ [$\mathrm{\hat{I}}$
is a 4$\times $4 unit matrix and $\left\langle u\left( \mathbf{r}\right)
u\left( \mathbf{r}^{\prime }\right) \right\rangle =u^{2}\delta \left( 
\mathbf{r}-\mathbf{r}^{\prime }\right) $] do not contain pseudospin
operators $\Lambda _{l}$, \textit{i.e.}, they remain invariant under the
group SU$_{2}^{\Lambda }$ transformations. Since $\vec{\Sigma}$ and $\vec{%
\Lambda}$ change sign under the time-inversion \cite{timereversal}, the
products $\Sigma _{s}\Lambda _{l}$ are $t\rightarrow -t$ invariant and,
together with $\mathrm{\hat{I}}$ can be used as a basis to represent
non-magnetic static disorder. Below, we assume that remote charges dominate
the elastic scattering rate, $\tau ^{-1}\approx \tau _{0}^{-1}\equiv \pi
\gamma u^{2}/\hbar $, where $\gamma =p_{\mathrm{F}}/(2\pi \hbar ^{2}v)$ is
the density of states of quasiparticles per spin in one valley. All other
types of disorder which originate from atomically sharp defects \cite%
{Ziegler} and break the SU$_{2}^{\Lambda }$ pseudospin symmetry are included
in a time-inversion-symmetric \cite{timereversal} random matrix $\Sigma
_{s}\Lambda _{l}u_{s,l}(\mathbf{r})$. Here, $u_{z,z}(\mathbf{r})$ describes
different on-site energies on the $A$ and $B$ sublattices. Terms with $%
u_{x,z}(\mathbf{r})$ and $u_{y,z}(\mathbf{r})$ take into account
fluctuations of $A\leftrightarrows B$ hopping, whereas $u_{s,x}(\mathbf{r})$
and $u_{s,y}(\mathbf{r})$\ generate inter-valley scattering. In addition,
warping term ${\hat{h}}_{\mathrm{w}}$ not only breaks $\mathbf{p}\rightarrow
-\mathbf{p}$ symmetry of the Fermi lines within each valley but also
partially lifts SU$_{2}^{\Lambda }$-symmetry.

Hidden SU$_{2}^{\Lambda }$ symmetry of the dominant part of ${\hat{H}}$ in
Eq. (\ref{h1-2}) enables us to classify the two-particle correlation
functions, 'Cooperons' which determine the interference correction to the
conductivity, $\delta g$ by pseudospin. Below, we show that $\delta g$ is
determined by the interplay of one pseudospin singlet ($C^{0}$) and three
triplet ($C^{x,y,z}$) Cooperons, $\delta g\propto -C^{0}+C^{z}+C^{x}+C^{y}$,
some of which are suppressed due to a lower symmetry of the Hamiltonian in
real graphene structures. That is, the 'warping' term ${\hat{h}}_{\mathrm{w}%
} $ and the disorder $\Sigma _{s}\Lambda _{z}u_{s,z}$ suppress intravalley
Cooperons $C^{x,y}$ and wash out the Berry phase effect and WAL, whereas
intervalley disorder $\Sigma _{s}\Lambda _{x(y)}u_{s,x(y)}(\mathbf{r})$
suppresses $C^{z}$ and restores weak localisation \cite{WL} of electrons,
provided that their phase coherence is long. This results in a WL-type
negative weak field magnetoresistance in graphene, which is absent when the
intervalley scattering time is long, as we discuss at the end of this Letter.

To describe quantum transport of 2D electrons in graphene we (a) evaluate
the disorder-averaged one-particle Green functions, vertex corrections,
Drude conductivity and transport time; \ (b) classify Cooperon modes and
derive equations for those which are gapless in the limit of purely
potential disorder; \ (c) analyse 'Hikami boxes' \cite{WL,LarkinWAL} for the
weak localisation diagrams paying attention to a peculiar form of the
current operator for Dirac electrons and evalute the interference correction
to conductivity leading to the WL magnetoresistance. In these calculations,
we treat trigonal warping ${\hat{h}}_{\mathrm{w}}$ in the free-electron
Hamiltonian Eqs. (\ref{h1},\ref{h1-2}) perturbatively, assume that potential
disorder $\mathrm{\hat{I}}u(\mathbf{r})$ dominates in the elastic scattering
rate, $\tau ^{-1}\approx \tau _{0}^{-1}=\pi \gamma u^{2}/\hbar $, and take
into account all other types of disorder when we determine the relaxation
spectra of low-gap Cooperons.

\textbf{(a).} Standard methods of the diagrammatic technique for disordered
systems \cite{WL,LarkinWAL} at $p_{\mathrm{F}}v\tau \gg \hbar $ yield the
disorder averaged single particle Green's function, 
\begin{equation}
\hat{G}^{R/A}\left( \mathbf{p},\epsilon \right) =\frac{\epsilon _{R/A}+v\,%
\vec{\Sigma}\mathbf{p}}{\epsilon _{R/A}^{2}-v^{2}p^{2}},\;\;\epsilon
_{R/A}=\epsilon \pm \tfrac{1}{2}i\hbar \tau _{0}^{-1}.  \notag
\end{equation}

\begin{figure}[t]
\centerline{\epsfxsize=\hsize\epsffile{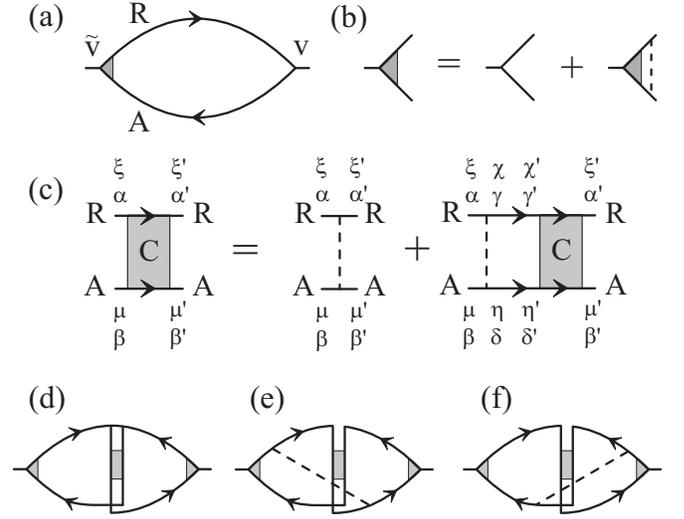}}
\caption{(a) Diagram for the Drude conductivity with (b) the vertex
correction. (c) Bethe-Salpeter equation for the Cooperon propagator with
valley indices $\protect\xi \protect\mu \protect\xi ^{\prime }\protect\mu %
^{\prime }$ and AB lattice indices $\protect\alpha \protect\beta \protect%
\alpha ^{\prime }\protect\beta ^{\prime }$. (d) Bare 'Hikami box' relating
the conductivity correction to the Cooperon propagator with (e) and (f)
dressed 'Hikami boxes'. Solid lines represent disorder averaged $G^{R/A}$,
dashed lines represent disorder.}
\label{fig:2}
\end{figure}

The current operator, $\mathbf{\hat{v}}=v\vec{\Sigma}$ for the Dirac-type
particles described in Eq. (\ref{h1}) is a momentum-independent. As a
result, the current vertex $\tilde{v}_{j}$ ( $j=x,y$), which enters the
Drude conductivity, Fig. 1(a), 
\begin{eqnarray}
g_{jj} &=&\frac{e^{2}}{\pi \hbar }\int \frac{d^{2}p}{\left( 2\pi \right) ^{2}%
}\mathrm{Tr}\left\{ \tilde{v}_{j}\hat{G}^{R}\left( \mathbf{p},\epsilon
\right) \hat{v}_{j}\hat{G}^{A}\left( \mathbf{p},\epsilon \right) \right\} , 
\notag \\
&=&4e^{2}\gamma D,\;\;\mathrm{with}\;\;D=v^{2}\tau _{0}\equiv \tfrac{1}{2}%
v^{2}\tau _{\mathrm{tr}},  \label{conductivity}
\end{eqnarray}%
is renormalised by vertex corrections in Fig.~\ref{fig:2}(b): $\mathbf{%
\tilde{v}}=2\mathbf{\hat{v}}=2v\vec{\Sigma}$. Here '$\mathrm{Tr}$' stands
for the trace over the AB and valley indices. The transport time in graphene
is twice the scatering time, $\tau _{\mathrm{tr}}=2\tau _{0}$, due to the
scattering anisotropy (lack of bacskattering off a potential scatterer).
This follows from the Einstein relation Eq. (\ref{conductivity}) (where spin
degeneracy has been taken into account).

%%%%%%%%%%%%%%%%% FIGURE 2 %%%%%%%%%%%%%

%%%%%%%%%%%%%%%%%%%%%%%%%%%%%%%%%%%%%%%%

\textbf{(b)}. The WL correction to the conductivity is associated with the
disorder-averaged two-particle correlation function $C_{\alpha \beta ,\alpha
^{\prime }\beta ^{\prime }}^{\xi \mu ,\xi ^{\prime }\mu ^{\prime }}$ known
as the Cooperon. It obeys the Bethe-Salpeter equation represented
diagrammatically in Fig.~\ref{fig:2}(c). The shaded blocks in Fig.~\ref%
{fig:2}(c) are infinite series of ladder diagrams, while the dashed lines
represent the correlator of the disorder in Eq. (\ref{h1-2}). Here, the
valley indices ($\mathbf{K}_{\pm }$) of the Dirac-type electron are included
as superscripts with incoming $\xi \mu $ and outgoing $\xi ^{\prime }\mu
^{\prime }$, and the sublattice ($AB$) indices as subscripts $\alpha \beta $
and $\alpha ^{\prime }\beta ^{\prime }$.

It is convenient to classify Cooperons in graphene as iso- and pseudospin
singlets and triplets, 
\begin{eqnarray}
C_{s_{1}s_{2}}^{l_{1}l_{2}} &=&\frac{1}{4}\sum_{\alpha ,\beta ,\alpha
^{\prime },\beta ^{\prime },}\sum_{\xi ,\mu ,\xi ^{\prime },\mu ^{\prime
},}\left( \Sigma _{y}\Sigma _{s_{1}}\Lambda _{y}\Lambda _{l_{1}}\right)
_{\alpha \beta }^{\xi \mu }  \notag \\
&&\quad \times \,C_{\alpha \beta ,\alpha ^{\prime }\beta ^{\prime }}^{\xi
\mu ,\xi ^{\prime }\mu ^{\prime }}\left( \Sigma _{s_{2}}\Sigma _{y}\Lambda
_{l_{2}}\Lambda _{y}\right) _{\beta ^{\prime }\alpha ^{\prime }}^{\mu
^{\prime }\xi ^{\prime }}\,.  \label{Cspin}
\end{eqnarray}%
Such a classification of modes is permitted by the commutation of the iso-
and pseudospin operators $\vec{\Sigma}$ and $\vec{\Lambda}$ in Eqs. (\ref%
{Sigma},\ref{Lamda},\ref{Cspin}), $[\Sigma _{s},\Lambda _{l}]=0$. To select
the isospin singlet ($s=0$) and triplet ($s=x,y,z$) Cooperon components
(scalar and vector representation of the group SU$_{2}^{\Sigma }\equiv
\{e^{ia\vec{n}\cdot \!\vec{\Sigma}}\}$), we project the incoming and
outgoing Cooperon indices onto matrices $\Sigma _{y}\Sigma _{s_{1}}$and $%
\Sigma _{s_{2}}\Sigma _{y}$, respectively. The pseudospin singlet ($l=0$)
and triplet ($l=x,y,z$) Cooperons (scalar and vector representation of the
'valley' group SU$_{2}^{\Lambda }\equiv \{\mathrm{e}^{ib\vec{n}\cdot \!\vec{%
\Lambda}}\}$) are determined by the projection of $C_{\alpha \beta ,\alpha
^{\prime }\beta ^{\prime }}^{\xi \mu ,\xi ^{\prime }\mu ^{\prime }}$ onto
matrices $\Lambda _{y}\Lambda _{l_{1}}$ ($\Lambda _{l_{2}}\Lambda _{y}$) and
are accounted for by superscript indices in $C_{s_{1}s_{2}}^{l_{1}l_{2}}$.

For disorder $\mathrm{\hat{I}}u(\mathbf{r})$, the equation in Fig. \ref%
{fig:2}(c) is 
\begin{multline}
C_{s_{1}s_{2}}^{l_{1}l_{2}}\left( \mathbf{q}\right) =\tau _{0}\,\delta
^{l_{1}l_{2}}\delta _{s_{1}s_{2}}  \notag \\
+\,\frac{1}{4\pi \gamma \tau _{0}\hbar }\sum_{s,l}C_{ss_{2}}^{ll_{2}}\left( 
\mathbf{q}\right) \int \frac{d^{2}p}{\left( 2\pi \right) ^{2}} \\
\times \,\mathrm{Tr}\Big\{\Sigma _{s}\Sigma _{y}\Lambda _{l}\Lambda _{y}%
\left[ \hat{G}_{\mathbf{p},\hbar \omega +\epsilon }^{R}\right] ^{\mathrm{t}%
}\Lambda _{y}\Lambda _{l_{1}}\Sigma _{y}\Sigma _{s_{1}}\hat{G}_{\hbar 
\mathbf{q}-\mathbf{p},\epsilon }^{A}\Big\}.  \notag
\end{multline}%
It leads to a series of coupled equations for the Cooperon matrix $\mathbf{C}%
^{l}$ with components $C_{ss^{\prime }}^{ll}$. It turn out that for
potential disorder~$\mathrm{\hat{I}}u(\mathbf{r})$ isospin-singlet modes $%
C_{00}^{ll}$ are gapless in all (singlet and triplet) pseudospin channels,
whereas triplet modes $C_{xx}^{ll}$ and $C_{yy}^{ll}$ have relaxation gaps $%
\Gamma _{x}^{l}=\Gamma _{y}^{l}=\frac{1}{2}\tau _{0}^{-1}$ and $C_{zz}^{ll}$
have gaps $\Gamma _{z}^{l}=\tau _{0}^{-1}$. When obtaining the diffusion
equations for the Cooperons using the gradient expansion of the
Bethe-Salpeter equation we take into account its matrix structure. The
matrix equation for each set of four Cooperons $\mathbf{C}^{l}$,where $%
l=0,x,y,z$ has the form%
\begin{equation*}
\left( 
\begin{array}{cccc}
\frac{1}{2}v^{2}\tau _{0}q^{2}+\Gamma _{0}^{l}-i\omega  & \frac{-i}{2}vq_{x}
& \frac{-i}{2}vq_{y} & 0 \\ 
\frac{-i}{2}vq_{x} & \frac{1}{2}\tau _{0}^{-1} & 0 & 0 \\ 
\frac{-i}{2}vq_{y} & 0 & \frac{1}{2}\tau _{0}^{-1} & 0 \\ 
0 & 0 & 0 & \tau _{0}^{-1}%
\end{array}%
\right) \mathbf{C}^{l}=\hat{1}.
\end{equation*}%
After the isospin-triplet modes were eliminated, the diffusion operator for
each of the four gapless/low-gap modes $C_{0}^{l}$ becomes $Dq^{2}-i\omega
+\Gamma _{0}^{l}$, where $D=\frac{1}{2}v^{2}\tau _{\mathrm{tr}}=v^{2}\tau
_{0}$.

Symmetry-breaking perturbations lead to relaxation gaps $\Gamma _{0}^{l}$ in
the otherwise gapless pseudospin-triplet components, $%
C_{0}^{x},C_{0}^{y},C_{0}^{z}$ of the isospin-singlet Cooperon $C_{0}^{l}$,
though they do not generate a relaxation of the pseudospin-singlet $%
C_{0}^{0} $ protected by the time-reversal symmetry of the Hamiltonian (\ref%
{h1-2}). We include all scattering mechanisms described in Eq. (\ref{h1-2})
in the corresponding disorder correlator (dashed line) on the r.h.s. of the
Bethe-Salpeter equation and in the scattering rate in the disorder-averaged $%
G^{R/A}$, as $\tau _{\mathrm{0}}^{-1}\rightarrow \tau ^{-1}=\tau _{\mathrm{0}%
}^{-1}+\sum_{sl}\tau _{sl}^{-1}$. For simplicity, we assume that different
types of disorder are uncorrelated, $\langle u_{s,l}(\mathbf{r})u_{s^{\prime
},l^{\prime }}(\mathbf{r}^{\prime })\rangle =u_{sl}^{2}\delta _{ss^{\prime
}}\delta _{ll^{\prime }}\delta (\mathbf{r}-\mathbf{r}^{\prime })$ and, on
average, isotropic in the $x-y$ plane: $u_{xl}^{2}=u_{yl}^{2}\equiv u_{\bot
l}^{2}$, $u_{sx}^{2}=u_{sy}^{2}\equiv u_{s\bot }^{2}$. We parametrize them
by scattering rates $\tau _{sl}^{-1}=\pi \gamma u_{sl}^{2}/\hbar $, where $%
\tau _{sx}^{-1}=\tau _{sy}^{-1}\equiv \tau _{s\bot }^{-1}$ and $\tau
_{xl}^{-1}=\tau _{yl}^{-1}\equiv \tau _{\bot l}^{-1}$ due to the $x-y$ plane
isotropy of disorder, which are combined into the intervalley scattering
rate $\tau _{\mathrm{i}}^{-1}$ and the intra-valley rate $\tau _{\mathrm{z}%
}^{-1}$, as 
\begin{equation}
\tau _{\mathrm{i}}^{-1}=4\tau _{\bot \bot }^{-1}+2\tau _{z\bot
}^{-1},\;\;\tau _{\mathrm{z}}^{-1}=4\tau _{\bot z}^{-1}+2\tau _{zz}^{-1}.
\label{iv}
\end{equation}

The trigonal warping term, ${\hat{h}}_{\mathrm{w}}$ in the Hamiltonian (\ref%
{h1}) plays a crucial role for the interference effects since it breaks the $%
\mathbf{p}\rightarrow -\mathbf{p}$ symmetry of the Fermi lines within each
valley: $\epsilon (\mathbf{K}_{\pm },-\mathbf{p})\neq \epsilon (\mathbf{K}%
_{\pm },\mathbf{p})$, while $\epsilon (\mathbf{K}_{\pm },-\mathbf{p}%
)=\epsilon (\mathbf{K}_{\mp },\mathbf{p})$ \cite{kpoints}. It has been
noticed \cite{EggShaped} that such a deformation of a Fermi line of 2D
electrons suppresses Cooperons.\ As ${\hat{h}}_{\mathrm{w}}$ has a similar
effect, it suppresses the pseudospin-triplet intravalley components $%
C_{0}^{x}$ and $C_{0}^{y}$, at the rate 
\begin{equation}
\tau _{\mathrm{w}}^{-1}=2\tau _{0}\left( \epsilon ^{2}\mu /\hbar
v^{2}\right) ^{2}.  \label{tauW}
\end{equation}%
However, since warping has an opposite effect on valleys $\mathbf{K}_{+}$
and $\mathbf{K}_{-}$, it does not cause gaps in the intervalley Cooperons $%
C_{0}^{0}$ (the only true gapless Cooperon mode) and $C_{0}^{z}$.

Altogether, the relaxation of modes $C_{0}^{l}$ can be described by the
following combinations of rates: 
\begin{equation*}
\Gamma _{0}^{0}=0,\;\Gamma _{0}^{z}=2\tau _{\mathrm{i}}^{-1},\;\Gamma
_{0}^{x}=\Gamma _{0}^{y}=\tau _{\mathrm{w}}^{-1}+\tau _{\mathrm{z}%
}^{-1}+\tau _{\mathrm{i}}^{-1}\equiv \tau _{\ast }^{-1}.
\end{equation*}%
In the presence of an external magnetic field, $\mathbf{B}=\mathrm{rot}%
\mathbf{A}$ and inelastic decoherence, $\tau _{\varphi }^{-1}$, equations
for $C_{0}^{l}\equiv $ $C_{00}^{ll}$ read 
\begin{equation*}
\lbrack D(i\mathbf{\nabla }+\tfrac{2e}{c\hbar }\mathbf{A)}^{2}+\Gamma
_{0}^{l}+\tau _{\varphi }^{-1}-i\omega ]C_{0}^{l}\left( \mathbf{r},\mathbf{\
r^{\prime }}\right) =\delta \left( \mathbf{r}-\mathbf{r^{\prime }}\right) .
\end{equation*}

\textbf{(c)}. Due to the momentum-independent form of the current operator $%
\mathbf{\tilde{v}=}2v\vec{\Sigma}$, the WL correction to conductivity $%
\delta g$ includes two additional diagrams, Fig.~\ref{fig:2}(e) and (f)
besides the standard diagram shown in Fig.~\ref{fig:2}(d). Each of the
diagrams in Fig.~\ref{fig:2}(e) and (f) [not included in the analysis in Ref.%
\cite{AndoWL}] produces a contribution equal to $(-\frac{1}{4})$ of that in
Fig.~\ref{fig:2}(d). This partial cancellation, together with a factor of
four from the vertex corrections and a factor of two from spin degeneracy
leads to 
\begin{equation}
\delta g=\frac{2e^{2}D}{\pi \hbar }\!\int \!\frac{d^{2}q}{\left( 2\pi
\right) ^{2}}\left( C_{0}^{x}+C_{0}^{y}+C_{0}^{z}-C_{0}^{0}\right) .
\label{Coop-WL}
\end{equation}

Using Eq. (\ref{Coop-WL}), we find the $B=0$ temperature dependent
correction, $\delta \rho $ to the graphene sheet resistance, 
\begin{equation}
\frac{\delta \rho \left( 0\right) }{\rho ^{2}}=-\delta g=\frac{e^{2}}{\pi h}%
\left[ \ln (1+2\frac{\tau _{\varphi }}{\tau _{\mathrm{i}}})-2\ln \frac{\tau
_{\varphi }/\tau _{\mathrm{tr}}}{1+\frac{\tau _{\varphi }}{\tau _{\ast }}}%
\right] ,  \label{WLZeroField}
\end{equation}%
and evaluate magnetoresistance, $\rho (B)-\rho (0)\equiv $ $\Delta \rho (B)$%
, 
\begin{gather}
\Delta \rho (T,B)=-\,\frac{e^{2}\rho ^{2}}{\pi h}\left[ F(\frac{B}{%
B_{\varphi }})-F(\frac{B}{B_{\varphi }+2B_{\mathrm{i}}})\right.   \notag \\
\;\;\;\;\;\;\;\;\;\;\;\;\;\;\;\;\;\;\;\;\;\;\left. -2F(\frac{B}{B_{\varphi
}+B_{\ast }})\right] ,  \label{Dsigma} \\
F(z)=\ln z+\psi (\frac{1}{2}+\frac{1}{z}),\;B_{\varphi ,\mathrm{i},\ast }=%
\frac{\hbar c}{4De}\tau _{\varphi ,\mathrm{i},\ast }^{-1}\,.  \notag
\end{gather}%
Here, $\psi $ is the digamma function, and the decoherence $\tau _{\varphi
}^{-1}(T)$ determines the MR curvature at $B\lesssim B_{\varphi }$.

%%%%%%%%% FIGURE 1 %%%%%%%%%%%%%%

\begin{figure}[t]
\centerline{\epsfxsize=\hsize\epsffile{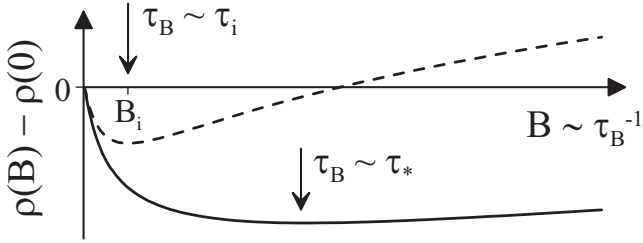}}
\caption{MR expected in a phase-coherent graphene $\protect\tau _{\protect%
\varphi }\gg \protect\tau _{\mathrm{i}}$: with $\protect\tau _{z},\protect%
\tau _{\mathrm{w}}\gg \protect\tau _{\mathrm{i}}$ (dashed) and $\protect\tau %
_{\ast }\ll \protect\tau _{\mathrm{i}}$ (solid line). In the case of $%
\protect\tau _{\protect\varphi }<\protect\tau _{\mathrm{i}}$, $\protect%
\delta \protect\rho =0$, so that $\Delta \protect\rho (B)=0$. }
\label{fig:1}
\end{figure}

%%%%%%%%%%%%%%%%%%%%%%%%%%%%%%%%%

Equations (\ref{Dsigma}) and (\ref{WLZeroField}) represent the main result
of this paper. They show that in graphene samples with the intervalley time
shorter than the decoherence time, $\tau _{\varphi }>\tau _{\mathrm{i}}$,
the quantum correction to the conductivity has the WL sign. Such behavior is
expected in graphene tightly coupled to the substrate (which generates
atomically sharp scatterers). Figure \ref{fig:1} illustrates the
corresponding MR in two regimes: $B_{\ast }\sim B_{\mathrm{i}}$ ($\tau
_{z},\tau _{\mathrm{w}}\gg \tau _{\mathrm{i}}$) and $B_{\ast }\gg B_{\mathrm{%
i}}$\ ($\tau _{\ast }\ll \tau _{\mathrm{i}}$). In both cases, the low-field
MR, at $B<B_{\mathrm{i}}$ is negative (for $B_{\ast }\sim B_{\mathrm{i}}$,
the MR changes sign at $B\sim B_{\mathrm{i}}$). A dashed line shows what one
would get upon neglecting the effect of warping, the solid curve shows the
MR behavior in graphene with a high carrier density, where the effect of
warping is strong and leads to a fast relaxation of intravalley Cooperons,
at the rate described in Eq. (\ref{tauW}). Then, in Eqs. (\ref{WLZeroField},%
\ref{Dsigma}) $\tau _{\ast }\approx \tau _{\mathrm{w}}\ll \tau _{\mathrm{i}%
}<\tau _{\varphi }$ and $B_{\ast }\gg B_{\mathrm{i}}$, which determines MR
of a distinctly WL type. Note that in the latter case MR is saturated at $%
B\sim B_{\mathrm{i}}$, in contrast to the WL MR in conventional electron
systems, where the logarithmic field dependence extends into the field range
of $\hbar c/4De\tau _{\mathrm{tr}}$. In a sheet loosely attached to a
substrate (or suspended), the intervalley scattering time may be longer than
the decoherence time, $\tau _{\mathrm{i}}>\tau _{\varphi }>\tau _{\mathrm{w}%
} $ ($B_{\mathrm{i}}<B_{\varphi }<B_{\mathrm{\ast }}$). In this case, $%
C_{0}^{z}$ in Eq. (\ref{Coop-WL}) is effectively gapless and cancels $%
C_{0}^{0}$, whereas trigonal warping suppresses the modes $C_{0}^{x}$ and $%
C_{0}^{y}$, so that $\delta g=0$ and MR displays neither WL nor WAL
behavior: $\Delta \rho (B)=0$.

Equation (\ref{Dsigma}) explains why in the recent experiments on the
quantum transport in graphene \cite{DiscGeim} the observed low-field MR
displayed a suppressed WL behavior rather than WAL. For all electron
densities in the samples studied in \cite{DiscGeim} the estimated
warping-induced relaxation time is rather short, $\tau _{\mathrm{w}}/\tau _{%
\mathrm{tr}}\sim 5\div 30$, $\tau _{\mathrm{w}}<\tau _{\varphi }$, which
excluded any WAL. Moreover, the observation \cite{DiscGeim} of a suppressed
WL MR in devices with a tighter coupling to the substrate agrees with the
behaviour expected in the case of sufficient intervalley scattering, $\tau _{%
\mathrm{i}}<\tau _{\varphi }$, whereas the absence of any WL MR, $\Delta
\rho (B)=0$ for a loosely coupled graphene sheet is what we predict for
samples with a long intervalley scattering time, $\tau _{\mathrm{i}}>\tau
_{\varphi }$.

In a narrow wire with the transverse diffusion time $L_{\bot }^{2}/D\ll \tau
_{\mathrm{i}},\tau _{\ast },\tau _{\varphi }$, edges scatter between valleys 
\cite{edge}. Thus, we estimate $\Gamma _{0}^{l}\sim \pi ^{2}D/L_{\bot }^{2}$
for the pseudospin triplet in a wire, whereas the singlet $C_{0}^{0}$
remains gapless. This yields negative magnetoresistivity for $B\lesssim 2\pi
B_{\bot }$, $B_{\bot }\equiv \hbar c/eL_{\bot }^{2}$: 
\begin{equation}
\frac{\Delta \rho _{\mathrm{wire}}\left( B\right) }{\rho ^{2}}=\frac{%
2e^{2}L_{\varphi }}{h}\,\left[ \frac{1}{\sqrt{1+\frac{1}{3}B^{2}/B_{\varphi
}B_{\bot }}}-1\right] .  \label{WL1D}
\end{equation}

Equations (\ref{WLZeroField}-\ref{WL1D}) completely describe the WL effect
in graphene and explain how the WL magnetoresistance reflects the degree of
valley symmetry breaking. They show that, despite the chiral nature of
electrons in graphene suggestive of antilocalisation, their long-range
propagation in a real disordered material or a narrow wire does not manifest
the chirality.

We thank I.Aleiner, V.Cheianov, A.Geim, P.Kim, O.Kashuba, and C.Marcus for
discussions. This project has been funded by the EPSRC grant EP/C511743.


\begin{thebibliography}{99}
\bibitem{kpoints} Here, $\mathbf{K}_{\pm }=\pm ({\textstyle\frac{2}{3}}%
ha^{-1},0)$, $a$ is the lattice constant. \
\bibitem{U4} The group U$_{4}$ can be described using 16 generators $\mathrm{%
\hat{I}},\Sigma _{s},\Lambda _{l},\Sigma _{s}\Lambda _{l}$; $s,l=x,y,x$.
\bibitem{timereversal} In the basis\ $\Phi =$[$\phi _{\mathbf{K}_{+},A}$, $%
\phi _{\mathbf{K}_{+},B}$, $\phi _{\mathbf{K}_{-,}B}$, $\phi _{\mathbf{K}%
_{-},A}$], time reversal of an operator $W$ is $T(\hat{W})=(\Pi _{x}\otimes
\sigma _{x}){W}^{\ast }(\Pi _{x}\otimes \sigma _{x})$, and $T(\Sigma
_{s})=-\Sigma _{s}$, $T(\Lambda _{l})=-\Lambda _{l}$, $T(\Sigma _{s}\Lambda
_{l})=\Sigma _{s}\Lambda _{l}$.
\end{thebibliography}
\end{document}